\documentclass[useAMS,usenatbib]{mn2e}
\usepackage{amsmath}
\usepackage{txfonts}
\usepackage{graphicx}
\usepackage{times}

\def\del#1{{}}

\newcommand{\hompc}{\,h\,{\rm Mpc}^{-1}}
\newcommand{\mpcoh}{\,h^{-1}\,{\rm Mpc}}

\newcommand{\apj}{ApJ}

\newcommand{\apjs}{ApJ Suppl.}
\newcommand{\aj}{Astron. J.}
\newcommand{\mnras}{MNRAS}

\begin{document}

\title[Scale dependent galaxy bias in the SDSS]
{Scale dependent galaxy bias in the SDSS as a function of \\luminosity and colour}

\author[James G. Cresswell, Will J. Percival]{
\parbox{\textwidth}{
James G. Cresswell\thanks{e-mail: jim.cresswell@port.ac.uk (JGC)} and
 Will J. Percival}
\vspace*{4pt} \\
Institute of Cosmology and Gravitation,\\ University of Portsmouth,
Portsmouth, PO1 2EG, UK}

\date{\today} 
\pagerange{\pageref{firstpage}--\pageref{lastpage}} \pubyear{2008}
\maketitle
\label{firstpage}

\begin{abstract}
  It has been known for a long time that the clustering of galaxies
  changes as a function of galaxy type. This galaxy bias acts as a
  hindrance to the extraction of cosmological information from the
  galaxy power spectrum or correlation function. Theoretical arguments
  show that a change in the amplitude of the clustering between
  galaxies and mass on large-scales is unavoidable, but cosmological
  information can be easily extracted from the shape of the power
  spectrum or correlation function if this bias is independent of
  scale. Scale-dependent bias is generally small on large scales,
  $k<0.1\hompc$, but on smaller scales can affect the recovery of
  $\Omega_mh$ from the measured shape of the clustering signal, and
  have a small effect on the Baryon Acoustic Oscillations. In this
  paper we investigate the transition from scale-independent to
  scale-dependent galaxy bias as a function of galaxy population. We
  use the Sloan Digital Sky Survey DR5 sample to fit various models,
  which attempt to parametrise the turn-off from scale-independent
  behaviour. For blue galaxies, we find that the strength of the
  turn-off is strongly dependent on galaxy luminosity, with stronger
  scale-dependent bias on larger scales for more luminous
  galaxies. For red galaxies, the scale-dependence is a weaker
  function of luminosity. Such trends need to be modelled in order to
  optimally extract the information available in future surveys, and
  can help with the design of such surveys.
\end{abstract}


\section{Introduction} \label{sec:intro}

Galaxies are not expected to form a Poisson sampling of the
distribution of matter in the Universe. Indeed, it has been known for
some time that different populations of galaxies demonstrate different
clustering strengths
\citep{davis76,dressler80,park94,pd94,seaborne99,norberg01,norberg02,zehavi02,zehavi05,li06},
showing that they cannot all have a simple relationship linking their
distribution with that of the matter. This galaxy bias severely limits
our ability to extract cosmological data from galaxy surveys
\citep{percival07,sanchez08}.

The large-scale shape of the linear matter power spectrum is dependent
on $\Omega_mh$ because of the change in the evolution of the Jeans
scale after matter-radiation equality
\citep{silk68,peebles70,sunyaev70,bond84,bond87,holtzman89}. However,
changes in the general shape of the power spectrum, such as that
caused by this physical process, are hard to separate from galaxy
bias, which also imprints a signature that changes smoothly with
scale. Galaxy bias can also affect mode-coupling in the power spectrum
by changing the non-linear scale, which can lead to small changes in
the Baryon Acoustic Oscillation positions
\citep{crocce08,matsubara08}. Such effects are beyond the scope of our
work, and we concentrate here on the broad changes to the shape of the
galaxy power spectrum.

The simplest model of galaxy bias is local, linear, deterministic
bias, $\delta_g({\bf x})=b_{\rm lin}\delta_{\rm lin}({\bf x})$ where
$\delta_g$ is the galaxy overdensity field, and $\delta_{\rm lin}$ is
the linear mass overdensity field. In this model, the bias $b_{\rm
  lin}$ is constant in space, but can change for different galaxy
populations. A more generic model may also include a stochastic
element which enters the description as an additional term,
$\delta_g({\bf x})=b_{\rm lin}\delta_{\rm lin}({\bf x})+\epsilon$. Or
the bias could be non-local, where it depends on a smoothed version of
the density field. In this paper, we will only be concerned with the
relation between the galaxy and mass power spectra, and therefore
define a practical measure of galaxy bias
\begin{equation}  \label{eq:bias}
  P_g(k) = b(k)^2P_{\rm lin}(k).
\end{equation}
Such a model can incorporate some of the complexities discussed above,
but is local in $k$-space, so cannot include any mode coupling terms,
which are expected to be present.

As we move to large scales, the galaxy bias is expected to tend
towards a constant value. The simplest model for this is the
peak-background split model \citep{BBKS,cole89}. Here galaxy formation
depends on the local density field. Large scale density modes can
alter the local galaxy number density by pushing pieces of the density
field above a critical threshold. On large scales we expect a linear
relationship between the large scale mode amplitudes and the change in
number density, so the shape of the galaxy and mass power spectra are
the same. Interestingly, such a linear relationship is broken if the
density field has a non-Gaussian component
\citep{dalal08,slosar08}. On small scales we expect galaxy clustering
to be different from that of the mass, as pairs of galaxies inside
single collapsed structures become important
\citep{seljak00,peacock00,cooray02}.

Although large-scale bias is seen as a nuisance when extracting
cosmological information from galaxy power spectra, it does provide
constraints on possible galaxy formation models. \citet{wild05}
proposed bivariate lognormal models of relative bias motivated by
observations of galaxy distributions. The data were found to support a
small (everywhere $<$5\%), but significant, amount of stochasticity
and non-linearity in these models on all scales, implying that galaxy
formation is not solely a function of local density; these effects
must be understood in order to properly utilise the next generation of
galaxy surveys. A related work, \citet{conway05}, also supported these
findings.

In order to use the power spectrum to extract cosmological
information, we need to investigate the transition between
scale-independent and scale-dependent galaxy bias. This transition is
known to be a function of the galaxy population chosen. Comparing work
in \citet{cole05} and \citet{percival07}, which used the same
techniques, and the analysis of \citet{sanchez08}, shows that there are
deviations between the shapes of the 2dFGRS and SDSS galaxy power
spectra. If this results from galaxy bias and the fact that the 2dFGRS
selected galaxies in a blue band whilst the SDSS selected galaxies in
a red band, then we should expect that similar changes show up in red
and blue galaxies drawn from a single survey. Using the SDSS,
\citet{percival07} showed that at $k=0.2\hompc$, the shape of the
power spectrum is a strong function of the $r$-band luminosity. In
this paper we extend this work by splitting in galaxy colour and
luminosity, and by fitting models to the resulting power spectra, in
order to see if the SDSS contains sufficient galaxies to fully explain
the trend observed between SDSS and 2dFGRS galaxies.

Previous work examining the observed dependence of the large scale
bias on luminosity was carried out by \citet{norberg01} who proposed
the phenomenological model $b=b_{1} + b_{2}L/L_{\star}$,
\citet{tegmark04} extended this to better fit the available data by
including an extra parameter in the form $b=b_{1} +
b_{2}L/L_{\star}+b_{3}(M-M_{\star})$.  \citet{swanson08} reexamined
both these models for samples split by galaxy colour. \citet{wild05}
also examined the colour dependence of relative bias. \citet{zehavi05}
presented colour dependent and luminosity dependent measurements of
the projected correlation function of SDSS galaxies. In all cases
subsets of the data existed where some aspect of the bias model in
question was shown to be a function of colour or luminosity.

We first describe the SDSS catalogues we use in
Section~\ref{sec:cats}, and how we model the radial selection function
of our subsamples using luminosity function fits in
Section~\ref{sec:sel_func}. Section~\ref{sec:calc_spectra} describes
our calculation of power spectra and uncertainties, and the bias
models to be fitted are presented in Section~\ref{sec:bias}. Results
of fitting the bias models are given in Section~\ref{sec:results}. In
Section~\ref{sec:bias_lum} we present simple models for the luminosity
dependence of our bias parameters and the results of fitting these to
our data. Conclusions and discussion are presented in
Section~\ref{sec:conclusions}.

\section{Galaxy Catalogues}
\label{sec:cats}

We use galaxy catalogues selected from the Sloan Digital Sky Survey
DR5 main galaxy sample. The Sloan Digital Sky Survey (SDSS;
\citealt{york00,adelman06}), which was recently completed, used a 2.5m
telescope \citep{gunn06} to obtain $10^4$ square degrees of imaging
data in five passbands $u$, $g$, $r$, $i$ and $z$
\citep{fukugita96,gunn98}. The main galaxy sample \citep{strauss02}
consists of galaxies with Petrosian $r$-band magnitude
$m_{r,petrosian} \leq 17.77$. This gives approximately $90$ galaxies
per square degree, with a median redshift $z=0.11$; in this paper we
use the DR5 sample \citep{adelman06}. We exclude a small subset of the
data taken during initial survey operation, for which the apparent
magnitude limit fluctuated. This gives $410\,095$ galaxies with $14.5
\leq m_{r,petrosian} \leq 17.77$ and redshift $>0.003$; the lower
redshift cut strongly reduces the contribution from mis-classified
stars. This is the catalogue used in \citet{percival07}, and further
details can be found here.

Where specified, we have K-corrected the galaxy luminosities using the
methodology outlined in \citet{blanton03a,blanton03b}. We also use the
same $z=0.1$ shifted $r$-band filter to define our luminosities (as
discussed in \citealt{blanton03b}), which we refer to as M$_{^{0.1}r}$
throughout this paper. Absolute magnitudes and $k$-corrections were
calculated assuming $h=100\,{\rm kms}^{-1}{\rm Mpc}^{-1}$,
$\Omega_M=0.3$ and $\Omega_\Lambda=0.7$, and we have applied the
recommended AB corrections to the observed SDSS magnitude system
\citep{smith02}.

\begin{table}
\begin{center}
\begin{tabular}{|c|c|c|c|}
bin    & absolute  & mean         & galaxy \\
       & magnitude & $M_{0.1_{r}}$ & count   \\
       & range     &              &        \\
\hline
red 1  & -22.30 ${\leq}$ $\,M_{0.1_{r}}<$ -21.35 & $-21.79\pm0.27$ & 49167\\
red 2  & -21.35 ${\leq}$ $\,M_{0.1_{r}}<$ -20.89 & $-21.12\pm0.13$ & 41462\\
red 3  & -20.89 ${\leq}$ $\,M_{0.1_{r}}<$ -20.47 & $-20.70\pm0.12$ & 37819\\
red 4  & -20.47 ${\leq}$ $\,M_{0.1_{r}}<$ -20.00 & $-20.27\pm0.14$ & 34651\\
red 5  & -20.00 ${\leq}$ \,$M_{0.1_{r}}<$ -19.34 & $-19.75\pm0.19$ & 29742\\
red 6  & -19.34 ${\leq}$ $\,M_{0.1_{r}}<$ -17.00 & $-19.01\pm0.38$ & 17582\\
       &                                        &                 &     \\
blue 1 & -22.30 ${\leq}$ $\,M_{0.1_{r}}<$ -21.35 & $-21.67\pm0.24$ & 17480\\
blue 2 & -21.35 ${\leq}$ $\,M_{0.1_{r}}<$ -20.89 & $-21.22\pm0.13$ & 25208\\
blue 3 & -20.89 ${\leq}$ $\,M_{0.1_{r}}<$ -20.47 & $-20.69\pm0.12$ & 28928\\
blue 4 & -20.47 ${\leq}$ $\,M_{0.1_{r}}<$ -20.00 & $-20.27\pm0.14$ & 32066\\
blue 5 & -20.00 ${\leq}$ \,$M_{0.1_{r}}<$ -19.34 & $-19.74\pm0.19$ & 36889\\
blue 6 & -19.34 ${\leq}$ $\,M_{0.1_{r}}<$ -17.00 & $-18.91\pm0.49$ & 48754\\
\end{tabular}
\end{center}
\caption{Description of the subcatalogues analysed in this paper. The limits of
the absolute magnitude bins are given, with the weighted mean and standard
deviation. We also give the galaxy count for each bin.}
\label{table:mag_bin_numbers}
\end{table}

\begin{figure}
\includegraphics[angle=0,width=\columnwidth, keepaspectratio=true]
  {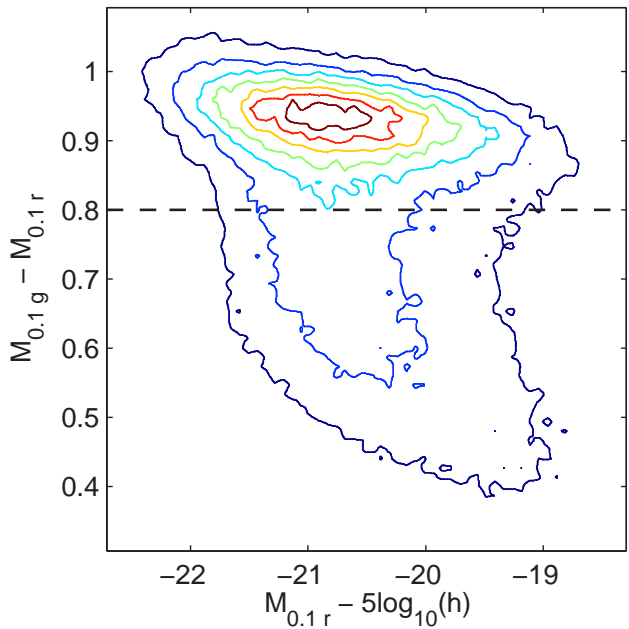}
  \caption{The distribution of galaxies in colour-absolute magnitude
    space. The dashed line denotes the boundary of the colour split at
    $M_{0.1_{g}} - M_{0.1_{r}} = 0.8$. \label{fig:ColourMagnitudeDiagram}}
\end{figure}

\begin{figure}
\includegraphics[angle=0,width=0.9\columnwidth, keepaspectratio=true]
{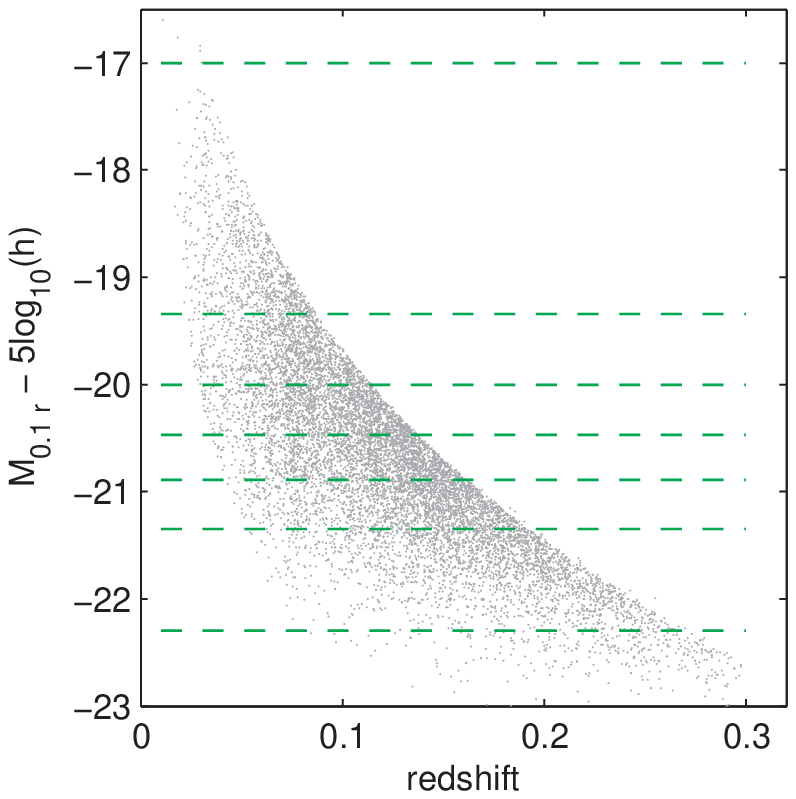}
\includegraphics[angle=0,width=0.9\columnwidth, keepaspectratio=true]
{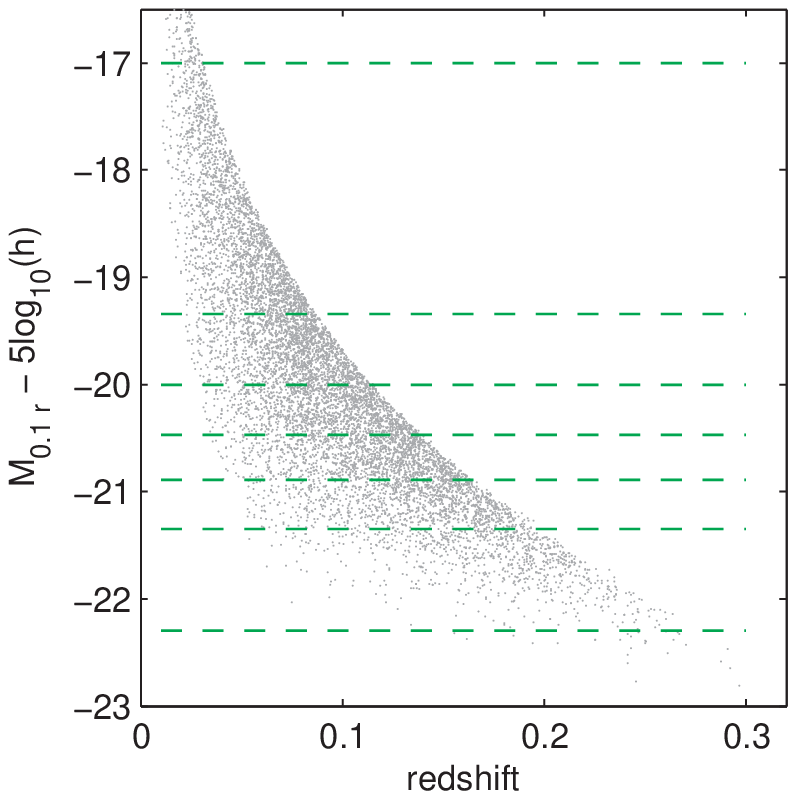}
\caption{The distribution of galaxies in absolute
magnitude--redshift space, red galaxies are shown in the top
panel, blue galaxies in the lower panel. The horizontal, dashed
lines denote the boundaries of the absolute magnitude
bins. \label{fig:MagnitudeRedshiftDiagram}}
\end{figure}

This sample of main galaxies was split into 6 absolute magnitude
limited subsamples giving approximately equal numbers of galaxies in
each bin. Each subsample was then further divided into red and blue
subsamples defined by a constant colour cut of
$M_{0.1_{g}}-M_{0.1_{r}} = 0.8$. Fig.~\ref{fig:ColourMagnitudeDiagram}
illustrates our chosen simple colour split, showing that this cut
clearly divides the two populations. We have tried more complicated
cuts without significant change in our results.

The absolute magnitude cuts and number of galaxies in each of these
subcatalogues is given in Table~\ref{table:mag_bin_numbers}.
Fig.~\ref{fig:MagnitudeRedshiftDiagram} illustrates the distribution
of red and blue galaxies within the imposed absolute magnitude
limits. At a given absolute magnitude the high and low redshift limits
result from the apparent magnitude limits of the survey, this can be
seen in Fig.~\ref{fig:MagnitudeRedshiftDiagram}. We could have cut
the catalogues in order to make these sub-catalogues volume limited,
but we wish to retain as much signal as possible. Significant cuts
would have been required in order to remove all effects of evolution
and K-corrections. \citet{percival07} showed a compromise using
``pseudo-volume limited'' catalogues. We choose instead to estimate
the redshift distribution by fitting the luminosity function (see the
next section), so we can model an apparent magnitude cut as easily as
a cut in absolute magnitude. This approach allows us to work with
considerably more galaxies and retain the maximum information.

\section{Modelling the Radial Selection Functions} 
\label{sec:sel_func}

For each of the catalogues described in the previous section, we need
to model the radial selection function. The angular mask is the same
for all catalogues as all the cuts being applied are independent of
angular position. We use a {\sc HEALPIX
  \footnote{http://healpix.jpl.nasa.gov }} \citep{gorski05} mask to
describe the angular galaxy distribution as described in
\citet{percival07}. We derive the radial distribution of galaxies in
each of our samples from fits to the luminosity functions for either
red or blue galaxies as appropriate. Section~\ref{subsec:LFs}
describes our luminosity function model and Section~\ref{subsec:f(z)}
describes our method for transforming this into a redshift
distribution.

\subsection{Redshift evolution corrected luminosity functions}
\label{subsec:LFs}

Our subsamples of red and blue galaxies contain sufficiently large
numbers of galaxies to allow us to calculate redshift evolution
corrected luminosity functions. We have found that well known
Schechter function \citep{schechter76} fits the data well. In terms of
absolute magnitude and with modifications to include redshift
evolution the Schechter function is given by
\begin{multline}
  \Phi(M,z)\,dM\,dV = \bar{n}\,0.4\log_{e}(10)\,10\,^{\,0.4\,(z-z\,_{0})\,P}\\
  10\,^{\,0.4\,(M^{\star}-M-Q\,(z-z\,_{0})\,)\,(1+\alpha)}\,
  exp\,{\{-10^{\,0.4\,(M^{\star}-M-Q\,(z-z\,_{0})\,)}\}}\,dM\,dV,
  \label{eqn:SchechterFunction}
\end{multline}
where $P$ and $Q$ are redshift evolution parameters following the
convention of \citet{lin99}: $P$ allows for density evolution, while
$Q$ allows for luminosity evolution. We have also tried fitting
non-parametric models, models with so many parameters that the final
shape of the function is to an extent independent of the shapes of the
contributing terms, based on the work of \citet{blanton03b}. However
we find no significant change in the resulting redshift distribution,
and we therefore only consider fitting a Schechter function in the
remainder of our paper.

\begin{figure}
\includegraphics[angle=0,width=\columnwidth, keepaspectratio=true]
  {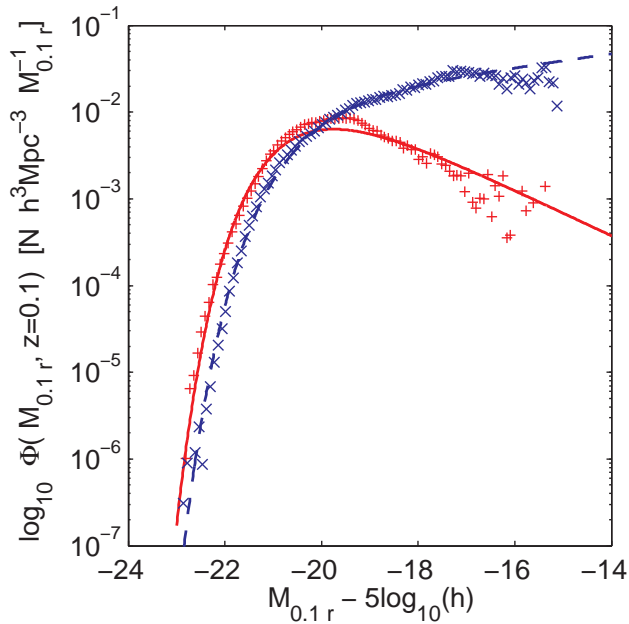}
  \caption{The best fit redshift-evolution corrected Schechter
    functions (Eq.\ref{eqn:SchechterFunction}, see text for details)
    representing the luminosity functions of the red galaxies, shown
    as the solid red line, and blue galaxies, shown as the dashed blue
    line, both at a redshift of $z=0.1$. The red, vertical crosses and
    blue, diagonal crosses represent the $1/V_{max}$ data estimates of
    the true luminosity function for red and blue galaxies
    respectively, redshift evolution corrected using the respective
    best fit parameters for Eq.~\ref{eqn:SchechterFunction} to a
    redshift of $z=0.1$. \label{fig:LF_Sch_RedVsBlue}}
\end{figure}
  
The luminosity function parameters were determined using a maximum
likelihood method \citep{lin99} implemented with the minimisation
routine Powell \citep{press92}. Convergence to the Likelihood maximum
was confirmed by starting the minimisation routine at a number of
widely separated initial parameter sets, and observing that the same
best-fit parameters were obtained. The results for the red and blue
galaxy samples can be seen in Fig.~\ref{fig:LF_Sch_RedVsBlue}. Note
that the overall normalisation, $\bar{n}$ cannot be determined by this
maximum likelihood method and we determine it directly from the
data. Further discussion of our procedure to calculate luminosity
functions and the parameters of the resulting fits will be presented
in \citet{cresswell08}. In this paper we simply use the luminosity
functions to estimate redshift distributions for the selected
catalogues, so the recovered parameters are not important, provided
the fits are an adequate match to the data: for instance, there is
some apparent discrepancy between the $1/V_{max}$ data estimates and
the best fit Schechter function for the red galaxies near an absolute
magnitude of $M_{0.1_{r}}\approx-19.3$; as the luminosity function is
slowly changing in this region and the normalisation is determined
independently of the fit this will have no significant effect on the
shape of any derived redshift distributions, which would here be
dominated by the effect of the apparent magnitude limits of the
survey.

\subsection{Redshift distributions}
\label{subsec:f(z)}

\begin{figure}
\includegraphics[angle=-0,width=\columnwidth, keepaspectratio=true]
  {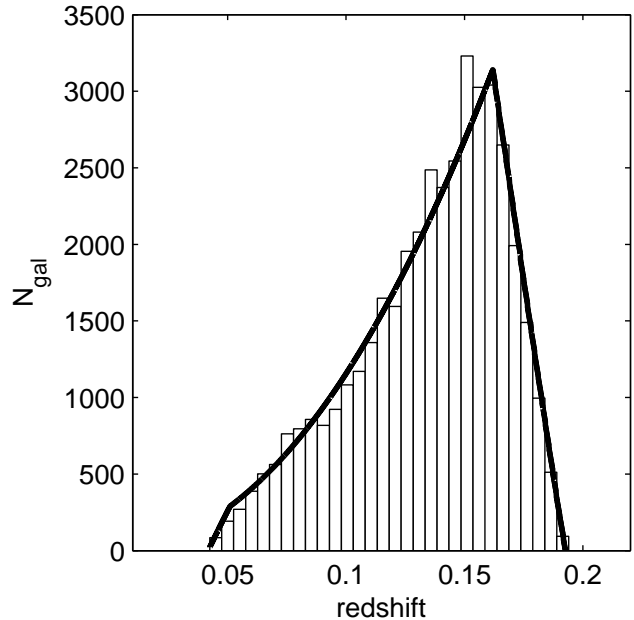}
  \caption{Histogram showing the redshift distribution of red galaxies
    in absolute magnitude bin 2. The thick black line is the model
    distribution derived from the red galaxy luminosity function via
    the relation shown in
    Eq.~(\ref{eqn:genericLfIntEq}). \label{fig:redshiftDists}}
\end{figure}

Given a luminosity function as defined in
Eq.~(\ref{eqn:SchechterFunction}) we can integrate to determine the
redshift distribution using
\begin{equation}  
  f(z)\,dz \propto \int_{M_{\rm lower}}^{M_{\rm upper}}
  \Phi(M^{\prime},z)dM^{\prime}\frac{dV}{dz}dz,  
  \label{eqn:genericLfIntEq}
\end{equation}
where $M_{\rm lower}$ is the minimum of the lower survey absolute
magnitude limit at a given redshift, and the lower bin magnitude
limit, and $M_{\rm upper}$ is similarly defined from the upper
boundaries of the sample and bin. An example of the data and model
redshift distributions for one of the subsamples, the red galaxies in
absolute magnitude bin 3, is shown in
Fig.~\ref{fig:redshiftDists}. Good agreement is seen in this plot, and
between model and data redshift distribution for all of our
subsamples, validating our procedure.

\section{calculating power spectra and uncertainties}
\label{sec:calc_spectra}

Power spectra were calculated as described in \citet{percival07},
using the standard Fourier technique of \citet{FKP}. The radial and
angular selection functions were included in this method by creating a
random catalogue with the same selection function as the galaxies, but
with Poisson sampling and 10$\times$ as many galaxies. The galaxies
were weighted using the optimal weights of \citet{FKP}. Obviously, no
additional bias-dependent weighting scheme (such as that in
\citealt{PVP} was applied).

For the SDSS DR5 sample we have created 2000 Log-Normal (LN)
catalogues, calculated as described in \citet{cole05}. We assumed a
flat $\Lambda$CDM power spectrum with $\Omega_mh=0.2$, and
$\Omega_b/\Omega_m=0.15$. Normalisation was matched to that of $L_*$ galaxies as
defined below. We have applied a colour and luminosity dependent bias
model that is scale-independent to these catalogues. The scheme was
iteratively matched to our results: initially we used the
luminosity-bias relation of \citet{norberg01} to calculate catalogues
and the corresponding covariance matrix for the data. Having fitted
the data, we then calculated new catalogues with a colour-luminosity
bias scheme that was a better fit to the data. This change resulted in
a negligible change to the best-fit bias model, so we are confident
that our results do not depend on this choice.

Power spectra were calculated from the mock catalogues using exactly
the same process as for the actual data, and the covariance matrices
were estimated for each of the subcatalogues described in
Section~\ref{sec:cats}. Because LN catalogues for each subcatalogue
were drawn from the same underlying density fields, we use these mocks
to calculate correlations between the power spectra for different
subcatalogues: these will be correlated as the volumes overlap.

\section{Fitting Models of Galaxy Bias}
\label{sec:bias}

As discussed in Section~\ref{sec:intro}, we define bias as the ratio
between the galaxy power spectra and the linear matter power spectrum
(Eq.~\ref{eq:bias}). We fit the observed power spectra with models
calculated from a linear model power spectra, using the fitting formulae of
\citet{eh98}, with parameters given by
the concordance cosmology $\Omega_0=0.241$, $\Omega_{\Lambda}=0.759$,
$H_0=73.2$, $\sigma_8=0.761$, $n_s=0.958$, $\Omega_b/\Omega_m=0.175$
\citep{spergel06}, multiplied by a bias model. The model is then
convolved with the window function for each catalogue, as described in
\citet{percival07}.  We will be interested in relatively large scales
$k<0.4\hompc$ and assume that, on these scales, the power spectrum
band-powers result from a multi-variate Gaussian distribution, and
that they are correlated within a single power spectrum, and between
different power spectra. We perform Maximum Likelihood fits to the
measured power spectra for three models of the bias $b(k)$. First we
assume that the bias is constant and fit on scales $k<0.21\hompc$. We
then consider two scale-dependent models for $b(k)$, fitting to
smaller scales. Results are given in section \ref{sec:results}.

First we use the $Q$-model \citep{cole05},
\begin{equation} \label{eqn:ScaleBias_eqn1} 
  b(k) = b_{\rm lin}\sqrt{\frac{1 + Qk^{2}}{ 1 + Ak}},
\end{equation}
where $b_{\rm lin}$ is the asymptotic large-scale bias, and $Q$ and
$A$ are parameters. \citet{cole05} used halo model catalogues to
suggest that $A=1.4$ in redshift-space leaving just a single nuisance
parameter $Q$ to be fitted to data. $Q$ would be expected to change
for different galaxy populations, and it is this that we test as a
function of galaxy colour and luminosity.

We also consider a model with
\begin{equation} \label{eqn:ScaleBias_eqn2}
  b(k) = b_{\rm lin}\sqrt{1+\frac{P}{b_{\rm lin}^2P_{\rm lin}}},
\end{equation}
where the parameter $P$ acts as an additional shot noise term. This
model has a physical basis as this term could account for a change in
the shot noise. This would arise if halos Poisson sampling the density
field, and galaxies are located in those halos, i.e. $P$ is the
contribution to the galaxy power spectrum from the one-halo term
\cite{seljak01,schulz06,guzik07}. We will refer to this model as the
$P$-model.

The relative merits of these two models are discussed in
\citet{smith06}, who argue in favour of the ease of physical
interpretation of the $P$-model. In addition, although they show that
for classes of cosmological model containing a free-streaming hot dark
matter component consisting of relic thermal axions, the $Q$-model
becomes highly pathological due to a degeneracy between $Q$ and the
particle mass. They suggest that this pathology may extend to other
models with light thermal relic components.

\section{Results from fitting the power spectra}
\label{sec:results}

\begin{figure}
\includegraphics[angle=0,width=\columnwidth, keepaspectratio=true]
  {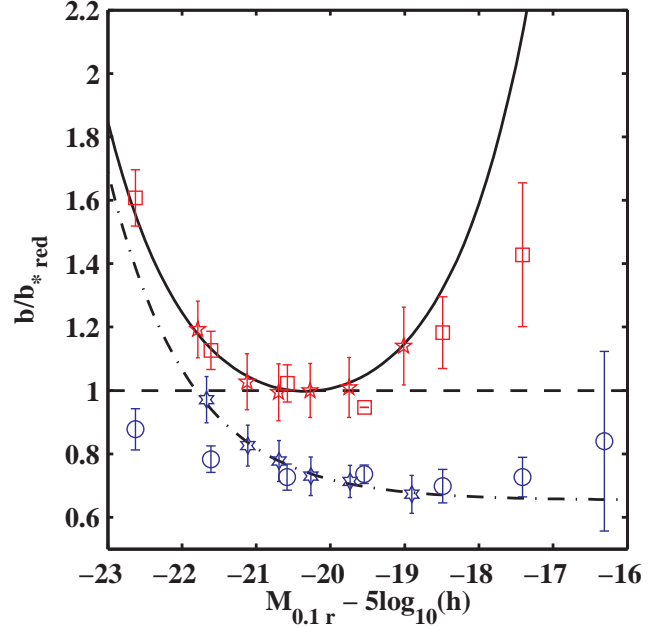}
  \caption{Comparison of relative, large-scale, constant, linear bias
    as a function of luminosity for galaxies split into red and blue
    colours. Five and six pointed stars represent our measurements of
    bias relative to our red galaxy $M_{\star}$ bin, for red and blue
    galaxies respectively. Squares and circles show the results from
    \citet{swanson08} renormalised to match our $M_{\star}$ values, see text for details. The upper, solid
    line is a fit to Eq.~\ref{eqn:LinearMagBias} for our red galaxy
    bias points, see text for details.  The lower, dash-dot line is
    the fit to blue galaxies. The horizontal dashed line shows no bias
    relative to our red $M_{\star}$ bin. \label{fig:relativeBias}}
\end{figure}

We have fitted bias models to power spectra calculated from the
catalogues parametrised in Table~\ref{table:mag_bin_numbers}. First we
consider fitting scale-independent bias on large scales, and then we
consider scale-dependent bias. All bias measurements were calculated
relative to the bias of red galaxies of luminosity $L_\star$,
$b_{\star,{\rm red}}$, where $L_\star$ was calculated from the fit to
the red-galaxy luminosity function.

\subsection{Large scale bias}
\label{subsec:relBias}

If we assume that the bias does not change with scale, and fit to very
large scales $k<0.21\hompc$, then the resulting bias amplitudes,
measured relative to $b_{\star,{\rm red}}$, are shown in
Fig.~\ref{fig:relativeBias}. For comparison we also plot the data of
\citet{swanson08} with an average evolution correction removed from
each magnitude bin. In order to compare samples with different
$L_\star$ values the \citet{swanson08} data have been offset to so
that the linear interpolation of the data points either side of our
$L_\star$ bin passes through our $L_\star$ bin, these corrections are
small, $\approx-0.05$ M$_{^{0.1}r}$ for red galaxies and
$\approx+0.01$ M$_{^{0.1}r}$ for blue galaxies. As can be seen, we
recover the same trends with the bias of blue galaxies monotonically
increasing with galaxy luminosity, while the red galaxies show more
complicated behaviour with increased bias for both high and low
luminosity galaxies. This comparison is discussed further in
Section~\ref{sec:conclusions}.

\subsection{Scale dependent bias}
\label{subsec:galBias}

The best-fit parameters for the $Q$-model and $P$-model fits are
presented in Table~\ref{table:param_fits}. These numbers were
calculated by fitting each of the 12 power spectra with the bias model
multiplied by the linear power spectrum.  The best-fit values and
their trends as a function of galaxy colour and luminosity are
analysed further in Section~\ref{sec:bias_lum}.

\begin{table}
\begin{center}
\begin{tabular}{ccccc}
       & \multicolumn{2}{c}{$Q$-model} &  \multicolumn{2}{c}{$P$-model} \\ 
bin    & $b_{lin}$    & $Q$           &  $b_{lin}$    & $P$           \\
\hline
red 1  & $1.40\pm0.02$ &  $9.45\pm0.70$  & $1.35\pm0.02$ & $512\pm48$ \\
red 2  & $1.22\pm0.03$ &  $9.24\pm0.83$  & $1.17\pm0.03$ & $375\pm40$ \\
red 3  & $1.17\pm0.03$ &  $9.44\pm0.94$  & $1.13\pm0.03$ & $351\pm42$ \\
red 4  & $1.20\pm0.04$ &  $7.72\pm0.93$  & $1.16\pm0.04$ & $277\pm46$ \\
red 5  & $1.25\pm0.05$ &  $7.26\pm1.03$  & $1.20\pm0.05$ & $272\pm51$ \\
red 6  & $1.44\pm0.08$ &  $8.03\pm1.45$  & $1.39\pm0.07$ & $419\pm80$ \\
       &             &               &             &          \\
blue 1 & $1.09\pm0.03$ & $13.74\pm1.29$  & $1.04\pm0.03$ & $541\pm48$ \\
blue 2 & $0.96\pm0.03$ &  $9.89\pm1.34$  & $0.92\pm0.03$ & $274\pm40$ \\
blue 3 & $0.94\pm0.04$ &  $7.74\pm1.43$  & $0.90\pm0.04$ & $177\pm43$ \\
blue 4 & $0.89\pm0.05$ &  $7.44\pm1.74$  & $0.86\pm0.05$ & $151\pm46$ \\
blue 5 & $0.91\pm0.07$ &  $4.64\pm1.78$  & $0.87\pm0.06$ & $66\pm51$ \\
blue 6 & $0.92\pm0.14$ &  $2.99\pm2.97$  & $0.87\pm0.13$ & $17\pm80$ \\
\end{tabular}
\end{center}
\caption{Maximum Likelihood parameters calculated by fitting the 
  $Q$-model and $P$-model for scale-dependent galaxy bias as given in 
  Eqns.~\ref{eqn:ScaleBias_eqn1}~\&~\ref{eqn:ScaleBias_eqn2}, to power
  spectra calculated for the catalogues described in 
  Section~\ref{sec:cats}. Best-fit values are presented together 
  with 1-$\sigma$ uncertainties.}
\label{table:param_fits}
\end{table}

\begin{figure*}
\includegraphics[angle=0,width=0.9\textwidth, keepaspectratio=true]
{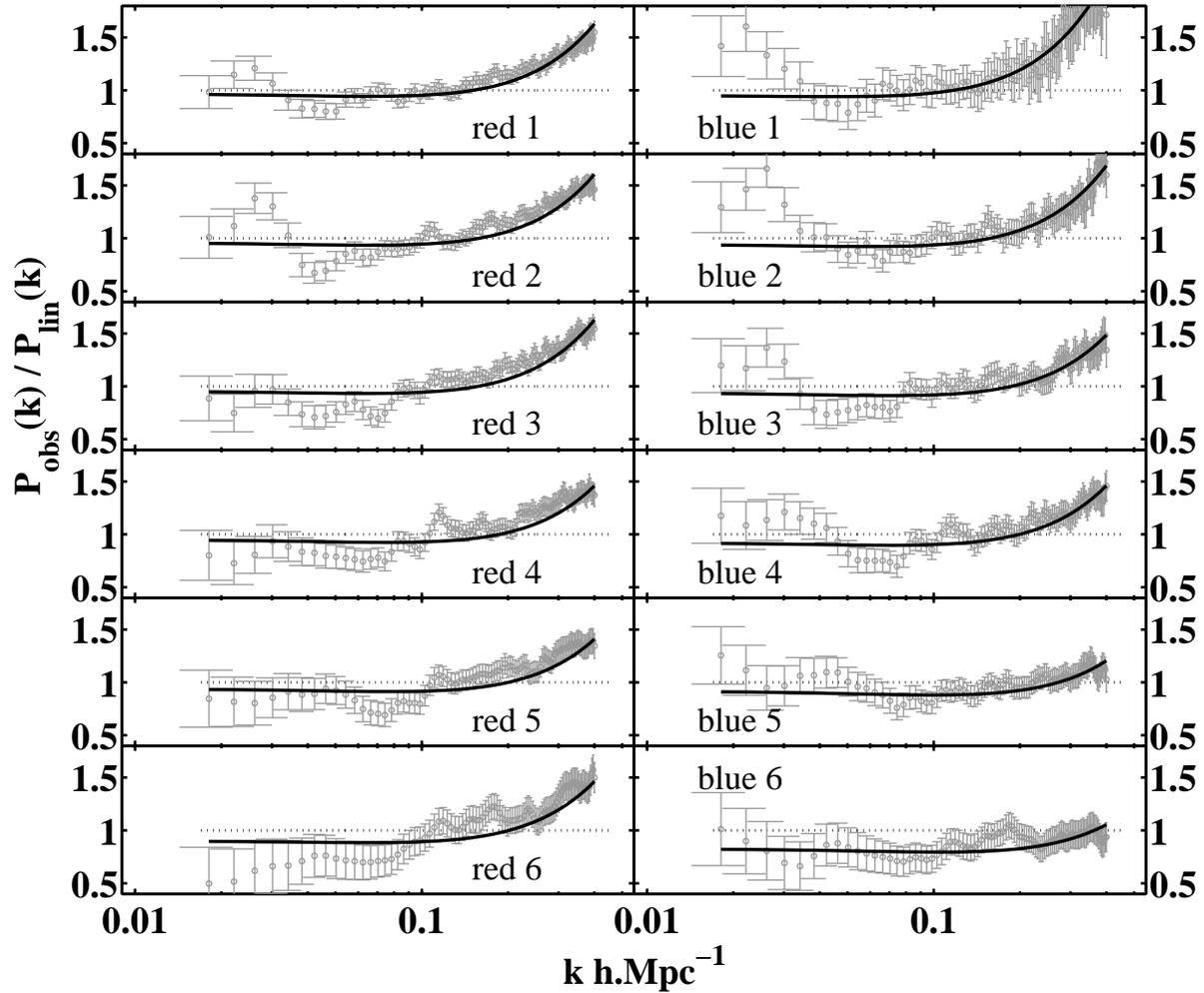}
\caption{Comparison of best-fit power spectra calculated with bias
  from the $Q$-model (solid lines) to the data (circles with 1$\sigma$
  errors). Each power spectrum is divided by our fiducial linear power
  spectrum convolved with the appropriate window
  function.\label{fig:Pk_Qmodel}}
\end{figure*}

\begin{figure}
\includegraphics[angle=0,width=\columnwidth, keepaspectratio=true]
{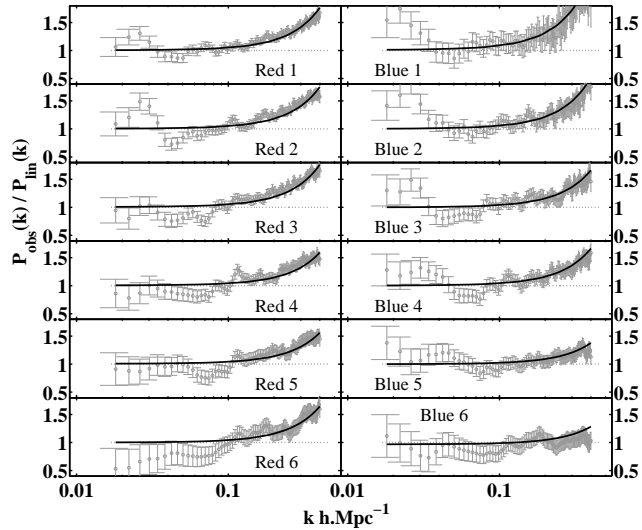}
\caption{As Fig.~\ref{fig:Pk_Qmodel}, but now for model power spectra
  calculated assuming the $P$-model for galaxy
  bias. \label{fig:Pk_Pmodel}}
\end{figure}

The model and data power spectra are compared in
Figs.~\ref{fig:Pk_Qmodel} \&~\ref{fig:Pk_Pmodel}. These data are
divided by our fiducial model convolved with the appropriate window
function. Comparisons are shown for the red galaxies in the left hand
panels and blue galaxies on the right, with the rows of panels showing
different absolute magnitude bins with the brightest at the top and
faintest at the bottom. The offset visible between the data and model
in the lower left panel is due to the high best-fit large-scale bias
in the best-fit solution (Eq.~\ref{eqn:LinearMagBias}, fit
parameters given in Table~\ref{table:param_fits}). Such effects arise
because the data are correlated, and the maximum Likelihood solution
does not match the $\chi$-by-eye expectation.

\section{Modelling bias as a function of luminosity}  
  \label{sec:bias_lum}

\begin{figure*}

\begin{center}
\includegraphics[angle=0,width=0.9\columnwidth, keepaspectratio=true]
  {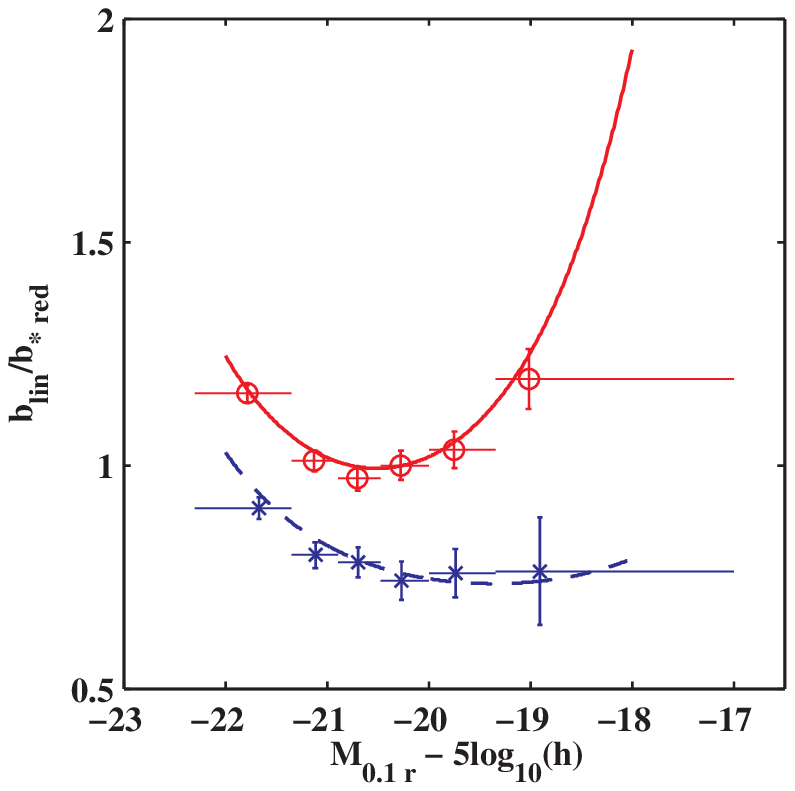}
\hspace{1cm}
\includegraphics[angle=0,width=0.9\columnwidth, keepaspectratio=true]
  {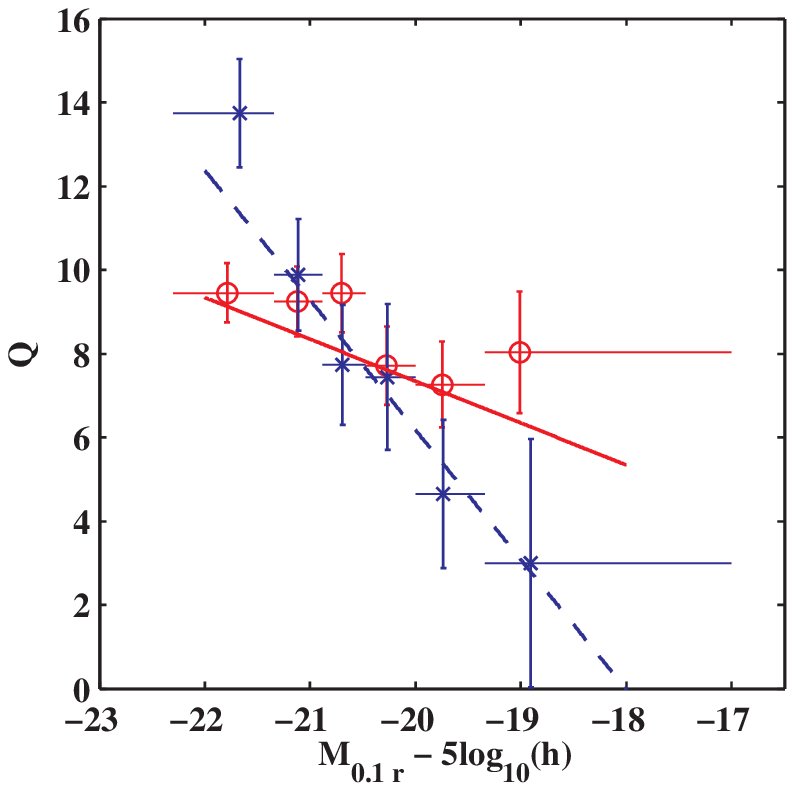}\\
\end{center}
\begin{center}
\includegraphics[angle=0,width=0.9\columnwidth, keepaspectratio=true]
  {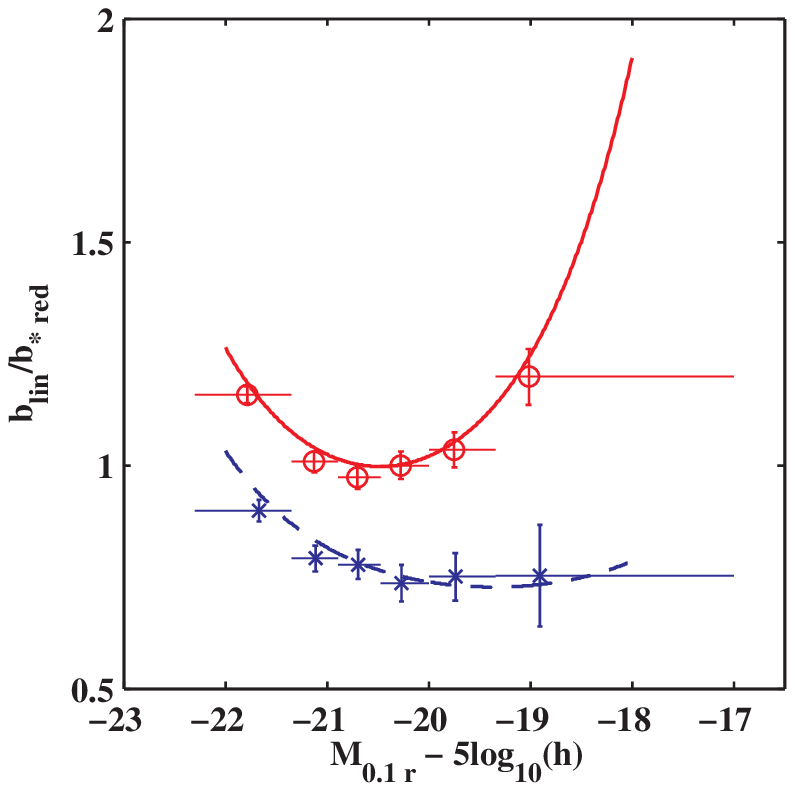}
\hspace{1cm}
\includegraphics[angle=0,width=0.9\columnwidth, keepaspectratio=true]
  {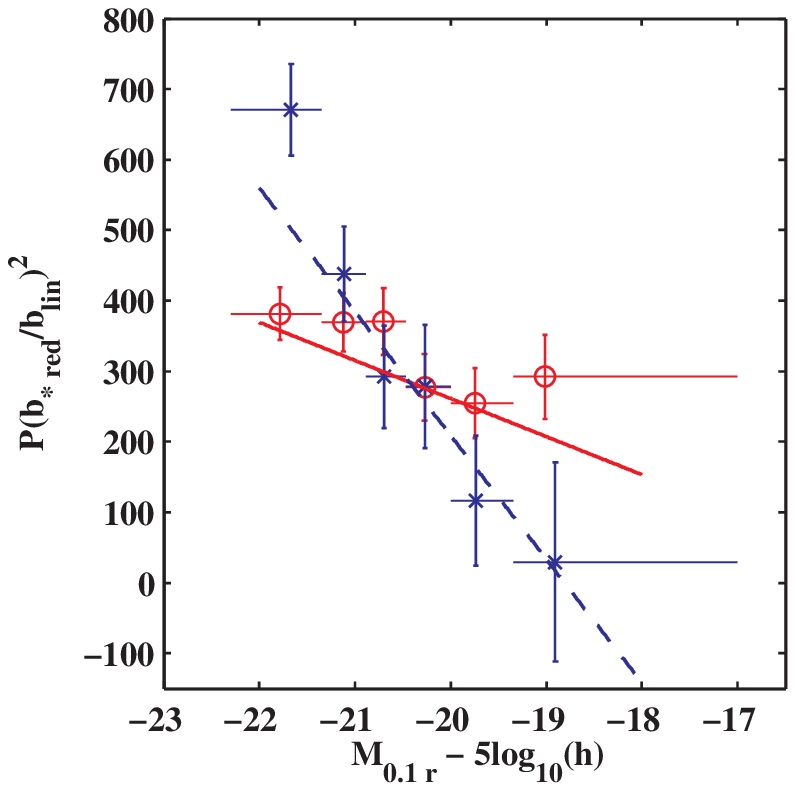}\\
\end{center}
\caption{Top row: best-fit $b_{lin}$, and $Q$ model parameters as a
  function of absolute magnitude, as presented in
  Table~\ref{table:param_fits}, from a fit including the $Q$-model
  bias prescription. Bottom row: best-fit $b_{lin}$, and $P$ model
  parameters as a function of absolute magnitude, as presented in
  Table~\ref{table:param_fits}, from a fit including the $P$-model
  bias prescription. Circles and crosses are for red and blue galaxies
  respectively. Solid, horizontal lines about each data point show the
  extent of each absolute magnitude bin. Solid lines show the models
  of Eqns.~\ref{eq:fit1_red}~\&~\ref{eq:fit2_red} and dashed lines
  show the models of ~\ref{eq:fit1_blue}~\&~\ref{eq:fit2_blue} in the
  appropriate panels for red and blue galaxies
  respectively. \label{fig:bias_param_vs_mag}}
\end{figure*}

In this section we try to fit a simple model for the luminosity
dependence of the parameters in the bias models given in
Eqns.~\ref{eqn:ScaleBias_eqn1}~\&~\ref{eqn:ScaleBias_eqn2}.

The asymptotic large-scale bias $b_{\rm lin}$ is known to be a
function of galaxy luminosity \citep{norberg01,tegmark04}, and galaxy
colour \citep{swanson08}. Here, we extend the form of the model
introduced by \citet{norberg01} in order to cope with the more
complicated behaviour seen when we also split galaxies by colour. We
assume the three parameter model
\begin{equation} \label{eqn:LinearMagBias}
  b_{\rm lin}(L) = a_1 +  a_2\frac{L}{L_\star} + a_3\frac{L_\star}{L}.
\end{equation}
For relative biases where $b_{\rm lin}(L_\star)=1$, we would have that
$a_1+a_2+a_3=1$, so the model would only have two free
parameters. This model reduces to the form of \citet{norberg01} in the
case of $a_3=0$.

In addition, we parametrise 
\begin{equation}
  Q(L) = q_1 + q_2(M-M_{\star}),
\end{equation}
and
\begin{equation}
  P(L) = p_1 + p_2(M-M_{\star}),
\end{equation}
where $q_1$, $q_2$, $p_1$ and $p_2$ are parameters that we can fit to
the data. We therefore have 5 free parameters in our fit to the data
in addition to choice of whether to use the P or $Q$-model. We have
fitted these parameters by performing a simultaneous Maximum
Likelihood search using all 12 power spectra, and allowing for
covariances between band powers in each, and between different power
spectra. I.e. we do not simply fit the recovered data values in
Fig.~\ref{fig:bias_param_vs_mag}, but instead perform a new fit to the
data. The resulting bias models are compared with the results
calculated when we allowed $b_{\rm lin}$, $Q$ and $P$ to match each
catalogue individually in Fig.~\ref{fig:bias_param_vs_mag}. As can be
seen, this simple model does very well in matching the
luminosity-dependent trends observed.

\begin{figure}
\includegraphics[angle=0,width=\columnwidth, keepaspectratio=true]
{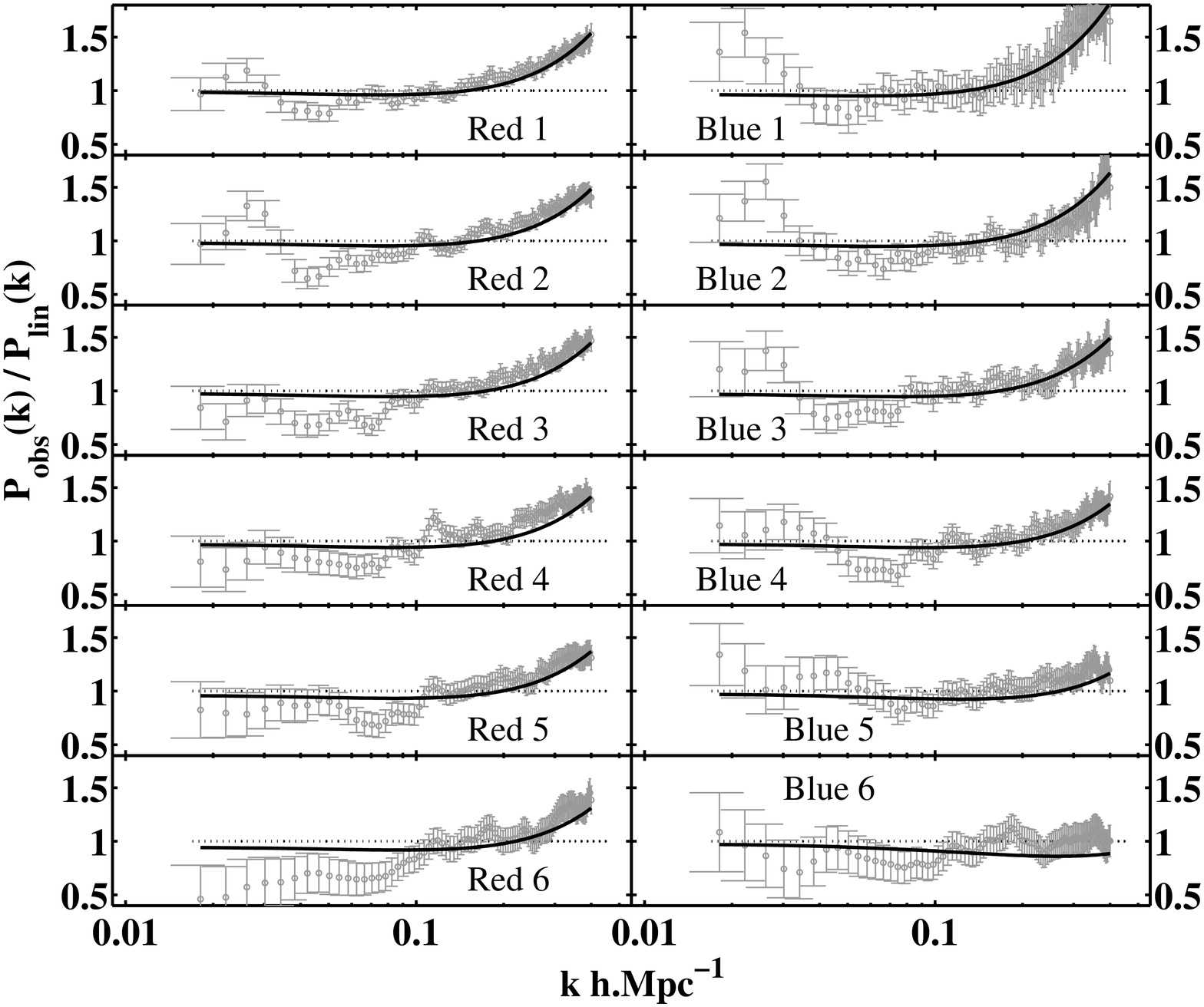}
\caption{As Fig.~\ref{fig:Pk_Qmodel}, but now showing models with
  parameters calculated from a simple fit to the luminosity dependent
  trends observed in
  Fig.~\ref{fig:bias_param_vs_mag}. \label{fig:Pk_Qmodel_bm}}
\end{figure}

The resulting best-fit parameters allow us to define the following
models for redshift-space galaxy power spectra. Using the $Q$-model, for
red galaxies,
\begin{multline} \label{eq:fit1_red}
P_{gal\, red}(L,k) = 
  \left(0.92+0.10\frac{L}{L_{\star}}+0.19\frac{L_{\star}}{L}\right) ^{2}\\
  \times P_{\rm lin}(L,k)\left(\frac{1+(7.5-1.0(M-M_{\star}))k^{2}}{1+Ak}\right),
\end{multline}
and for blue galaxies,
\begin{multline}  \label{eq:fit1_blue}
P_{gal\, blue}(L,k) = 
  \left(0.81+0.07\frac{L}{L_{\star}}+0.02\frac{L_{\star}}{L}\right) ^{2}\\
  \times P_{\rm lin}(L,k)\left(\frac{1+(6.3-3.1(M-M_{\star}))k^{2}}{1+Ak}\right).
\end{multline}
These model power spectra are compared with those observed in
Fig.~\ref{fig:Pk_Qmodel_bm}, where good agreement is seen.

\begin{figure}
\includegraphics[angle=0,width=\columnwidth, keepaspectratio=true]
{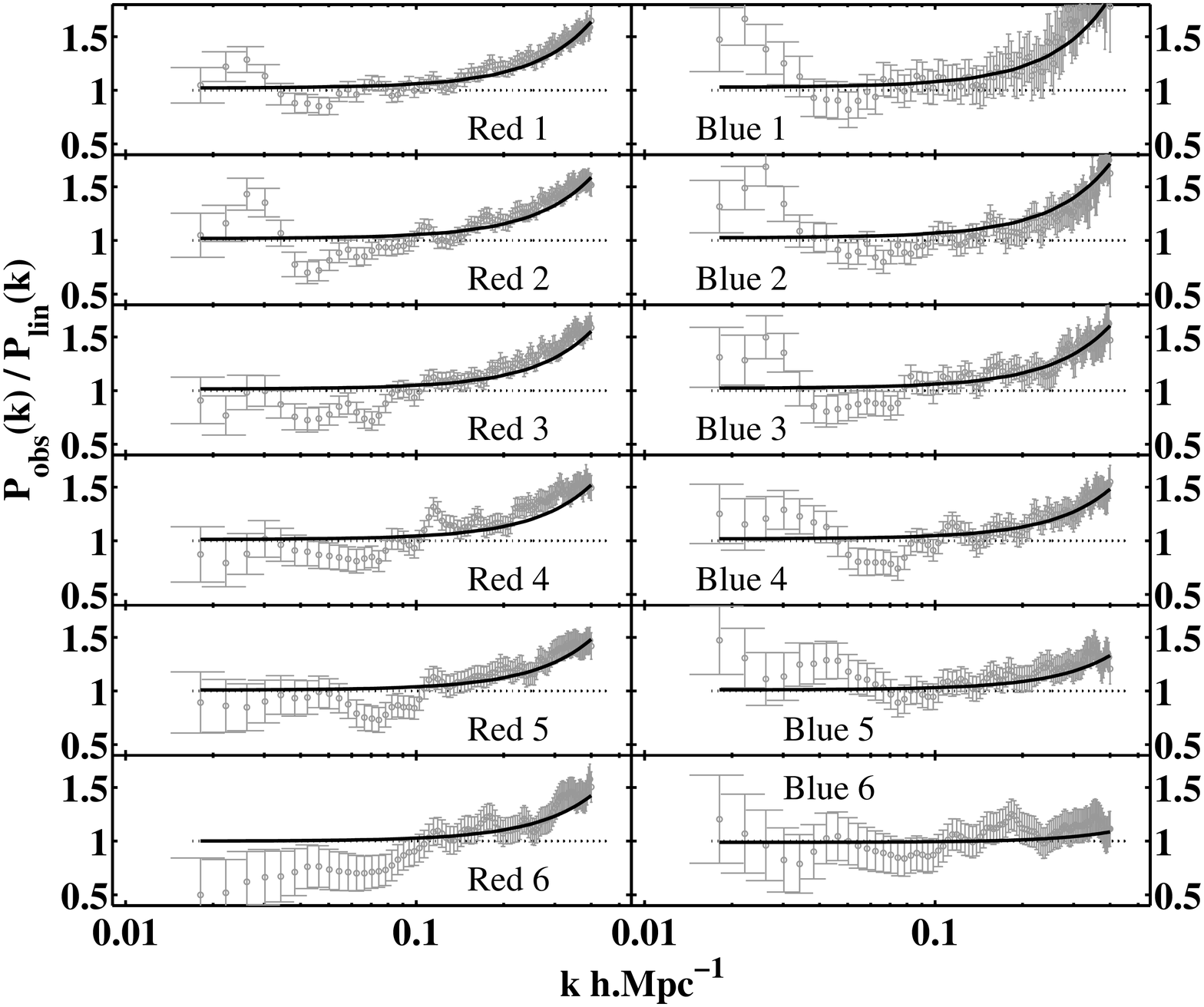}
\caption{As Fig.~\ref{fig:Pk_Pmodel}, but now showing models with
  parameters calculated from our simple fit to the luminosity
  dependent trends observed in
  Fig.~\ref{fig:bias_param_vs_mag}. \label{fig:Pk_Pmodel_bm}}
\end{figure}

Using the $P$-model, for red galaxies,
\begin{multline} \label{eq:fit2_red}
P_{gal\,red}(L,k) = 
  \left(0.89+0.10\frac{L}{L_{\star}}+0.18\frac{L_{\star}}{L}\right)^{2}\\
  \times{P_{\rm lin}(L,k)+\left(200-40(M-M_{\star})\right)},
\end{multline}
and for blue galaxies,
\begin{multline}  \label{eq:fit2_blue}
P_{gal\,blue}(L,k) = 
  \left(0.77+0.07\frac{L}{L_{\star}}+0.02\frac{L_{\star}}{L}\right)^{2}\\
  \times{P_{\rm lin}(L,k)+\left(160-130(M-M_{\star})\right)},
\end{multline}
where, as previously noted in the text, $A=1.4$. These model power spectra are
compared with those observed in
Fig.~\ref{fig:Pk_Pmodel_bm} where, as in Fig.~\ref{fig:Pk_Qmodel_bm},
good agreement is seen.

The $\chi^2$ values of the best-fit models including the $P$-model of
bias are good, with $609$ given $571$ degrees-of-freedom for the red
galaxies and $439$ for the blue galaxies given the same number of
degrees of freedom for the blue galaxies. For the $Q$-model, the
corresponding numbers are $624$, and $440$. These numbers depend
strongly on the covariance model adopted and, as a consequence, we do
not analyse these further other than to note that the fits seem to
give reasonable numbers.

\section{Conclusions}  \label{sec:conclusions}

We have quantified the bias of red and blue galaxies as a function of
luminosity on scales $k<0.4\hompc$. We find clear differences in the
large-scale asymptotic bias between blue and red galaxies, similar to
that found by \citet{zehavi05} and \citet{swanson08}, and shown in
Fig.~\ref{fig:bias_param_vs_mag}. At large scales red galaxies have
are more biased than blue galaxies for all luminosities. The bias of
blue galaxies is a strong, monotonically increasing function of
luminosity, with more luminous galaxies being the most biased. For red
galaxies, the picture is more complicated, with an increase in the
bias of faint red galaxies (\citet{zehavi05} find a similar result but
stress the scale dependence of the observation). This trend is
significant at the $7.2-\sigma$ level.We find a stronger bias for
luminous blue galaxies, compared with that of \citet{swanson08}. The
reason for this is unknown, although it is worth noting that our
catalogue is larger than that used by \citet{swanson08}, as they used
additional cuts, constructing volume limited subsamples, compared with
our sample selection procedure (see section \ref{sec:cats}). We
performed a simple test of the dependence of this difference on
observed scale, repeating our analysis for scales
$0.038<k<0.070\hompc$ which approximates the \citet{swanson08}
measurement scale of $\sim20\mpcoh$, the discrepancy in the bias of
the brightest blue galaxies was not removed. However, the change of
scales brought the best fit red galaxy linear bias model into better
agreement with the \citet{swanson08} faint, red galaxies; the model is
poorly constrained at the faint end due to the lack of unique data
points in that range. It is worth noting that, due to different colour
cut criteria, the brightest absolute magnitude bins in
\citet{swanson08} are inherently more red than ours, this might lead
one to expect larger bias measurements for those bins, the opposite of
the observed trend. The bright blue sample contains the smallest
numbers of galaxies, so the errors on the relative bias should be the
largest of any sample. However, our expected errors are insufficient
to fully explain the discrepancy, and there remains no obvious reason
for this difference.

The work on constant bias models has been extended by considering how
the turn-off from constant large-scale bias depends on galaxy colour
and luminosity. We have compared two models for this turn-off: the
$P$-model and $Q$-model given in Eqns.~\ref{eqn:ScaleBias_eqn1}
\&~\ref{eqn:ScaleBias_eqn2} respectively. We find that there is little
to choose between the two in terms of how well they can fit the
current data, and both provide adequate fits to the power spectrum
trends observed as a function of galaxy luminosity and colour at the
current level of data precision. Although there is no observational
motivation, \citet{hamann08} argued that the $P$-model has a physical
motivation, and therefore offers a more attractive model of bias.

We use the values of $P$ and $Q$ obtained for these two models to
quantify the degree of divergence from a constant bias model. Note
that these parameters change both the position and amplitude of the
turn-off. We find that the best-fit values of $P$ and $Q$, shown in
Fig.~\ref{fig:bias_param_vs_mag}, are a strong function of luminosity
for blue galaxies. Red galaxies show far weaker evolution with
luminosity, and are consistent with the hypothesis of no change in the
scale at which the bias can no longer be described as a constant, to
current data precision. Interestingly, this trend in the turn-off from
apparent scale-independent behaviour does not match that of the
amplitude of the large-scale bias. If amplitude and turn-off scale
were linked, we might have expected the $Q$ and $P$ parameters to
match for red galaxies of high or low luminosity, where the constant
bias component matches, but be different for intermediate luminosity
galaxies. We do not observe such a trend, suggesting that there might
not be a simple link between the amplitude of bias and the scale at
which the 1-halo term becomes important.

It is clear that the simple $P$ and $Q$-models will become
insufficient to model the observational data, both on very small
scales, and as the data improve. They are, after all, simply motivated
to fit observed trends, and do not encompass all of the physics
involved. One alternative and more complex prescription for bias
prediction is that presented in \cite{yoo08} which can predict
$P_{obs}(k)$ given a set of cosmological parameters and a HOD model,
the parameters of which have been determined from the observed,
projected correlation function (the correlation function provides
additional information to that of the power spectrum as they are
sensitive to different scales). However, the potential limits of
simple, phenomenological models, such as the $P$ and $Q$-models, does
not stop us being able to use them to investigate trends in the data,
as we have done in this paper. The current data clearly show that, on
relatively large scales (our small scale limit of $0.4\hompc$
approximately corresponds to $17\mpcoh$), blue and red galaxies have
very different bias properties. It is the blue galaxies that are faint
in the $r$-band whose relative clustering on different scales most
closely resembles that of the mass, and the luminous blue galaxies the
least. This supports the hypothesis that the shapes of the 2dFGRS and
SDSS main galaxy power spectra of \citet{cole05} and
\citet{tegmark04,percival07} differ because the average galaxy bias of
each sample differs, caused by different sample selections. The
clustering of the 2dFGRS galaxies, on average, would be expected to be
a better tracer of the linear clustering signal out to smaller scales
than the SDSS galaxies.

This highlights the importance of sample selection for future galaxy
surveys, and the importance of understanding galaxy bias for
extracting cosmological information from power spectrum shapes from
such surveys. Red and blue galaxies show very different trends in
their large-scale bias and the scale-dependent smaller-scale bias as a
function of luminosity. It is clear that galaxies need to be split
into sub-populations by more than just the luminosity in a single band
in order to properly understand and model bias.

\section*{Acknowledgements}

JC would like to thank Steven Peter Bamford and Bjoern Malte Schaefer
for helpful and insightful discussions. We thank Molly Swanson for
providing her data for comparison with ours in
Fig.~\ref{fig:relativeBias}, and for helpful discussions. We thank the
referee for helpful comments. JC is funded by a STFC PhD
studentship. WJP is supported by STFC, the Leverhulme Trust and the
European Research Council.

\label{lastpage}

\end{document}